\documentclass[twocolumn,showpacs,preprintnumbers,amsmath,amssymb,superscriptaddress]{revtex4}
\pdfoutput=1

\usepackage[pdftex]{graphicx}
\usepackage[pdftex,colorlinks=true,urlcolor=blue,linkcolor=black]{hyperref} 

\usepackage{graphicx}
\usepackage{dcolumn}
\usepackage{bm}
\usepackage{datetime}
\usepackage       			{ulem}
\usepackage{float}

\begin{document}

\title{Pairing in few-fermion systems with attractive interactions } 
\author{G.\,Z\"urn}
\affiliation{Physikalisches Institut, Ruprecht-Karls-Universit\"at Heidelberg, 69120 Heidelberg, Germany}
\affiliation{Max-Planck-Institut f\" ur Kernphysik, Saupfercheckweg 1, 69117 Heidelberg, Germany}
 \email{gerhard.zuern@physi.uni-heidelberg.de}
 \author{A.\,N.\,Wenz}
\affiliation{Physikalisches Institut, Ruprecht-Karls-Universit\"at Heidelberg, 69120 Heidelberg, Germany}
\affiliation{Max-Planck-Institut f\" ur Kernphysik, Saupfercheckweg 1, 69117 Heidelberg, Germany}
 \author{S.\,Murmann}
\affiliation{Physikalisches Institut, Ruprecht-Karls-Universit\"at Heidelberg, 69120 Heidelberg, Germany}
\affiliation{Max-Planck-Institut f\" ur Kernphysik, Saupfercheckweg 1, 69117 Heidelberg, Germany}
\author{A.\,Bergschneider}
\affiliation{Physikalisches Institut, Ruprecht-Karls-Universit\"at Heidelberg, 69120 Heidelberg, Germany}
\affiliation{Max-Planck-Institut f\" ur Kernphysik, Saupfercheckweg 1, 69117 Heidelberg, Germany}
 \author{T.\,Lompe}
\affiliation{Physikalisches Institut, Ruprecht-Karls-Universit\"at Heidelberg, 69120 Heidelberg, Germany}
\affiliation{Max-Planck-Institut f\" ur Kernphysik, Saupfercheckweg 1, 69117 Heidelberg, Germany}
\affiliation{ExtreMe Matter Institute EMMI, GSI Helmholtzzentrum f\"ur Schwerionenforschung, 64291 Darmstadt, Germany}
\author{S.\,Jochim}
\affiliation{Physikalisches Institut, Ruprecht-Karls-Universit\"at Heidelberg, 69120 Heidelberg, Germany}
\affiliation{Max-Planck-Institut f\" ur Kernphysik, Saupfercheckweg 1, 69117 Heidelberg, Germany}
\affiliation{ExtreMe Matter Institute EMMI, GSI Helmholtzzentrum f\"ur Schwerionenforschung, 64291 Darmstadt, Germany}

\date{\today} 

\begin{abstract}
We have studied
quasi one-dimensional few-particle systems consisting of one to six ultracold fermionic atoms in two different spin states with attractive interactions. We probe the system by deforming the trapping potential and by observing the tunneling of particles out of the trap. For even particle numbers we observe a tunneling behavior which deviates from uncorrelated single-particle tunneling indicating the existence of pair correlations in the system. From the tunneling timescales we infer the differences in interaction energies of systems with different number of particles which show a strong odd-even effect, similar to the one observed for neutron separation experiments in nuclei.

\end{abstract}

\maketitle

Pairing between distinguishable fermions with an attractive interparticle interaction leads to fascinating phenomena in a variety of vastly different systems. In metals at sufficiently low temperature pairs of electrons can form a superfluid, as described by Bardeen, Cooper and Schrieffer in their BCS-theory of superconductivity \cite{BCS1957}. Using dilute gases of ultracold atoms, where the interparticle interactions can be tuned freely using Feshbach resonances \cite{Chin2010} it was shown that such BCS pairs can be smoothly converted into bosonic molecules \cite{regal2003}, which leads to a continuous crossover from a BCS-like superfluid to a BEC of molecules \cite{leggett1980, nozieres1985, bartenstein2004, zwierlein2005}.  
In finite Fermi systems pairing has been studied extensively in the context of nuclear physics \cite{Migdal1959, Zelevinsky2003, Brink2005}. Here the pairing caused by the attractive interaction between the nucleons leads to an enhanced stability of systems with an even number of neutrons or protons \cite{Brink2005}. For systems with fully closed shells -- the so-called magic nuclei -- stability is further enhanced. 

Recently, it has become possible to prepare finite systems of ultracold fermions in well-defined quantum states \cite{Serwane2011}. In such a system one has direct experimental control over key parameters such as the particle number and the depth and shape of the confining potential. Combined with the ability to tune the interparticle interactions \cite{Chin2010, zuern2013}, this makes this system uniquely suited to study pairing in a controlled environment.

In this work we study how pairing affects few-particle systems consisting of one to six ultracold  atoms in two different spin states -- labeled $\mid\uparrow\rangle$ and $\mid \downarrow\rangle$ -- confined in a cigar-shaped optical microtrap \cite{hyperfinestate}. We deterministically prepare these systems in their ground state 
using the preparation scheme developed in \cite{Serwane2011}.
Our microtrap has typical trap frequencies of $\omega_{\parallel}=2 \pi \times 1.488(14)$\,kHz \cite{zuern2012} in longitudinal and $\omega_{\perp}=2 \pi \times 14.22(35)$\,kHz \cite{sala2013} in perpendicular direction. In addition to the optical potential we can apply a linear potential in longitudinal direction by applying a magnetic field gradient. A full description of the potential shape as determined in \cite{zuern2012} is given in \cite{SOM}.

\begin{figure} [tb]
\centering
	\includegraphics [width= 8.7cm] {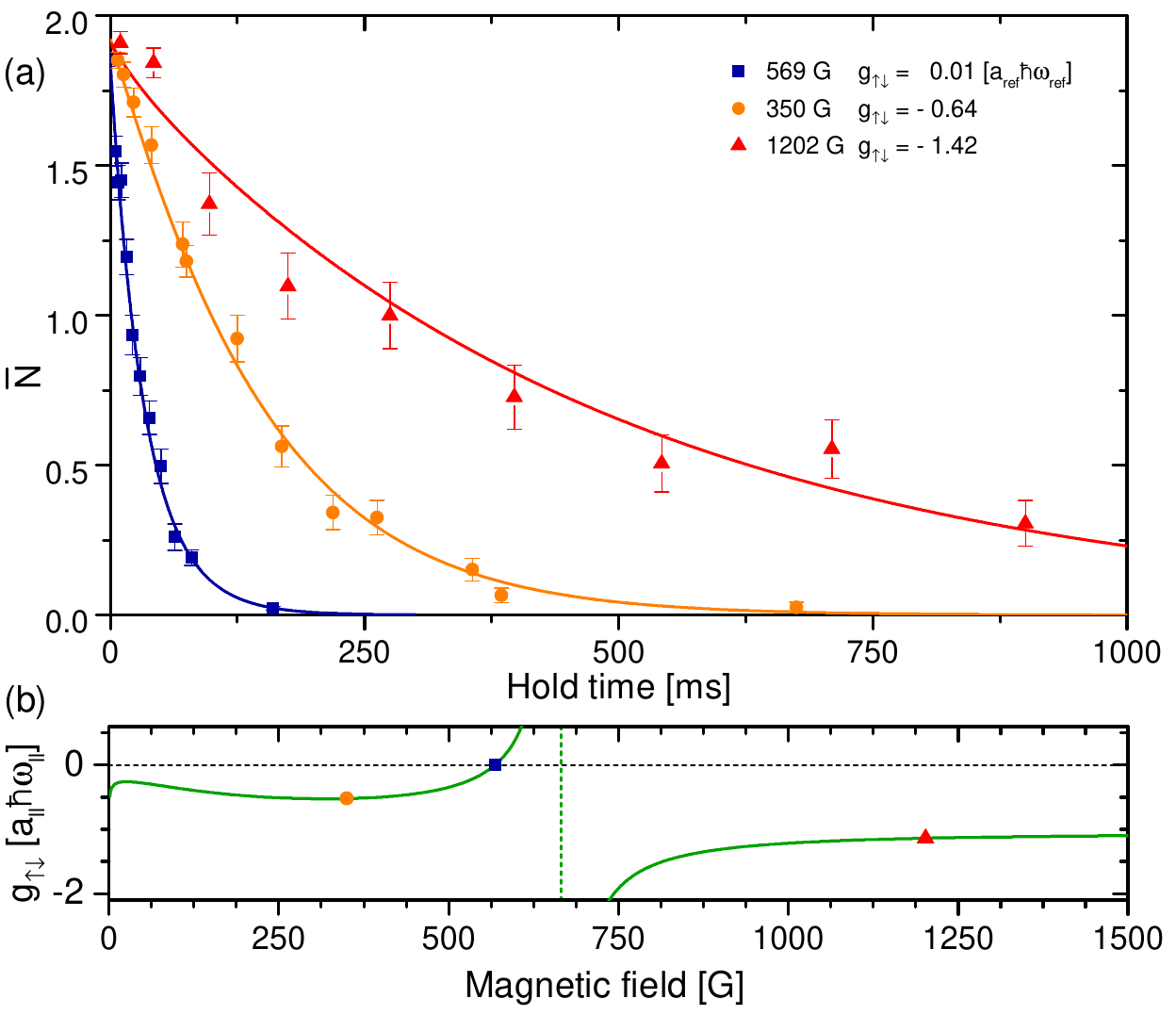}
\caption{To study a system of two attractively interacting fermions we study its tunneling dynamics by creating a barrier of fixed height and recording the number of particles remaining in the trap after different hold times. We find that the tunneling slows down as we increase the strength of the interparticle interaction from zero (blue squares) to intermediate (orange dots) or to larger values (red triangles), where the solid lines are fits according to the tunneling model described in the text and the errors are the standard error of the mean of $100$ individual measurements. This shows the increase of the effective barrier height due to the energy shift caused by the attractive interaction. 
(b) Coupling constant $g_{\uparrow \downarrow}$ in the untilted potential as a function of the magnetic field with $a_{\parallel}=\sqrt{\hbar/ \mu \omega_{\parallel}}$. Note that $g_{\uparrow \downarrow}$ depends on the parameters of the deformed potential and is therefore given in units of $a_{\text{ref}} \hbar \omega_{\text{ref}}$ for the different measurements \cite{SOM}.
}
	\label{fig:figure1}
\end{figure}
As in our few-fermion systems all energy scales are much smaller than $\hbar \omega_{\perp}$ the system can be treated as quasi one-dimensional \cite{Idziaszek2004}.
In this 1D environment the interaction between distinguishable particles can be described by a contact interaction whose coupling constant $g_{\uparrow \downarrow}$ can be tuned by a confinement induced resonance  \cite{Olshanii1998} (see Fig.\ref{fig:figure1}b).

In a first set of experiments we study the emergence of pair correlations in  a two-particle system. Therefore
we prepare two particles, one in state $\mid\uparrow\rangle$ and one in state $\mid \downarrow\rangle$, in the ground state of the trapping potential. To probe the system we employ the same method as described in \cite{zuern2012}:  We lower the depth of the optical potential such that there is a potential barrier of well-defined height through which the particles can tunnel out of the trap.  After a certain hold time we ramp the potential back up and measure the number of particles remaining in the trap. By performing many of these measurements at different hold times we measure the time evolution of the probabilities $P_2(t)$, $P_1(t)$ and $P_0(t)$ to find two, one or zero particles in the tilted potential. From these probabilities we get the mean particle number 
	$\overline{N} \left(t \right) = 2\,P_2\left(t \right) + 1\, P_1\left(t \right) \;$
whose time evolution is shown in Fig.\ref{fig:figure1}a) for three different values of the interparticle interaction.

For a system of two noninteracting particles the loss follows an exponential decay with a tunneling rate ${\gamma_{s_0}\approx30\,\frac{1}{\text{s}}}$. In the presence of an attractive interparticle interaction (i.e. $g_{\uparrow \downarrow} < 0$) the energy of the system is reduced. This leads to an effective increase in the height of the tunneling barrier and therefore the tunneling slows down. Consequently the tunneling of the particles is no longer independent and thus cannot be described by a simple exponential decay.
 
To describe the correlated tunneling of the two particles we use a simple model which takes into account two different loss processes (see inset in Fig.\,\ref{fig:figure2}). 
The first is pair tunneling, which we define as two particles leaving the trap at the same time. The rate at which this process occurs is labeled $\gamma_p$. The second process is subsequent single-particle tunneling. Here one particle tunnels first, while the other particle remains in the unperturbed ground state of the trap. In this case the first particle tunnels with a rate $2\gamma_s$ which is determined by the effective height of the tunneling barrier which in turn depends on the interaction energy of the two particles. For the second particle there is no interaction shift and consequently it leaves the trap with the rate $\gamma_{s_0}$ measured for the noninteracting system. 

To relate these rates to our measured probabilities $P_i(t)$ we set up a set of rate equations which give the probabilities to find two, one or zero particles in the trap as a function of the hold time. The probability $P_2(t)$ to find two particles in the trap decreases with the sum of the single-particle tunneling rate $2 \gamma_{s}$ and the pair tunneling rate $\gamma_{p}$:
\begin{equation}
	\frac{dP_2\left(t \right)}{dt}  = -(2 \gamma_s + \gamma_p) P_2\left(t \right) \; .
\label{eq:P2_rate_equation_normal}
\end{equation}
This rate equation can be easily solved and the decay law for the two particle probability reads:
\begin{equation}
	P_2\left(t \right)  = e^{-(2\gamma_s+\gamma_p )t}  \; .
\label{eq:P2_prob}
\end{equation}

The rate equation for the probability $P_1(t)$ is given by  
\begin{equation}
	\frac{dP_1\left(t \right)}{dt}  = 2 \gamma_s P_2\left(t \right) -\gamma_{s_0} P_1\left(t \right), 
\label{eq:P1_rate_equation_normal}
\end{equation}
where two-particle systems become one-particle systems with the rate 
$2 \gamma_{s}$, where $\gamma_{s}$ is the rate with which one of the particles leaves the trap. These systems
then decay into zero-particle systems with the rate $\gamma_{s_0}$.
Assuming a perfect preparation fidelity for the initial sample the initial conditions for this equation are $P_2(0)=1$ and $P_1(0)=P_0(0)=0$ and one obtains a probability 
\begin{equation}
	P_1\left(t \right)  = \frac{2 \gamma_s}{2\gamma_s+\gamma_p-\gamma_{s_0}}\left[ e^{-\gamma_{s_0}t}-e^{-\left(2 \gamma_s+\gamma_p \right)t} \right] \;.
\label{eq:P1_prob}
\end{equation}
to find a single particle in the trap.

To describe our experiments we have to take into account two additional effects. The first is the finite preparation fidelity $0 \leq f \leq 1$ which defines the starting conditions of the decay. The second is that changing the magnetic offset field to tune the interaction strength affects the magnetic moment of the atoms. This leads to a state-dependent change in the shape of the tilted potential as a function of the magnetic field \cite{SOM}. Consequently the tunneling rates also obtain a spin dependence which leads to the modified solutions
\begin{equation}
	P_2\left(t \right)  = f \, e^{-(\gamma_{s\vert \downarrow \rangle}+\gamma_{s\vert \uparrow \rangle}+\gamma_p)t} =f \, e^{-\gamma_2 t} 
\label{eq:P2_rate_equation_mod}
\end{equation}
and\\
\bgroup
\def\arraystretch{2}%
  \begin{tabular}{llr}
$\; \; \; \; \; \;  P_1\left( t \right)  =$&$f(\frac{\gamma_{s\vert \uparrow\rangle}}{\gamma_2-\gamma_{s_0\vert \downarrow \rangle}}    \left[ e^{-\gamma_{s_0\vert \downarrow \rangle}t}-e^{-\gamma_2 t}\right]$& \\
&$+\: \frac{\gamma_{s\vert \downarrow \rangle}}{\gamma_2-\gamma_{s_0\vert \uparrow \rangle}}    \left[ e^{-\gamma_{s_0\vert \uparrow \rangle}t}-e^{-\gamma_2 t}\right])$& \; \; \; \; \; (6) \\
	& $+(1-f)(\frac{1}{2}e^{-\gamma_{s_0\vert \uparrow \rangle}t} + \frac{1}{2} e^{-\gamma_{s_0\vert \downarrow \rangle}t})$&   \\
    \end{tabular}
with the spin dependent rate constants ${\gamma_{s\vert \uparrow\rangle}}$,  ${\gamma_{s\vert \downarrow\rangle}}$,  ${\gamma_{s_0\vert \uparrow\rangle}}$ and ${\gamma_{s_0\vert \downarrow\rangle}}$. In eq.\,(6) the first (second) term corresponds to a two-particle system where the $\mid \uparrow\rangle$ ($\mid \downarrow\rangle$) particle has tunneled first while the last term accounts for the single-particle systems which are present due to the finite preparation fidelity.

To determine the rates for single-particle and pair tunneling we first fit our data for $P_2\left(t \right)$ with eq.\,(\ref{eq:P2_rate_equation_mod}) and extract the combined loss rate $\gamma_2$ and the preparation fidelity. As fitting all remaining parameters to our data for $P_1(t)$ would be unstable we independently determine ${\gamma_{s_0\vert \uparrow\rangle}}$ and ${\gamma_{s_0\vert \downarrow\rangle}}$ in a series of separate experiments with single $\mid \uparrow\rangle$ or $\mid \downarrow\rangle$ particles \cite{SOM}. To further reduce the number of fit parameters we make use of the fact that ${\gamma_{s\vert \uparrow\rangle}}$ and ${\gamma_{s\vert \downarrow\rangle}}$ are linked by the shape of the potential which we can infer from our measurements of ${\gamma_{s_0\vert \uparrow\rangle}}$ and ${\gamma_{s_0\vert \downarrow\rangle}}$. 
 As we already know $\gamma_2 = \gamma_{s\vert \uparrow \rangle}+\gamma_{s\vert \downarrow \rangle}+\gamma_p$ this leaves only one free parameter for our fit: The pair tunneling rate $\gamma_p $.
Details on the fitting procedure can be found in \cite{SOM}.
As an example the measured values for $P_2(t)$, $P_1(t)$ and the resulting fits for an interaction strength of $g_{\uparrow \downarrow} = -0.64$ are shown in 	Fig.\ref{fig:figure2} \cite{unitg1d}.\\
\begin{figure} [tb!]
\centering
	\includegraphics [width=8.7cm] {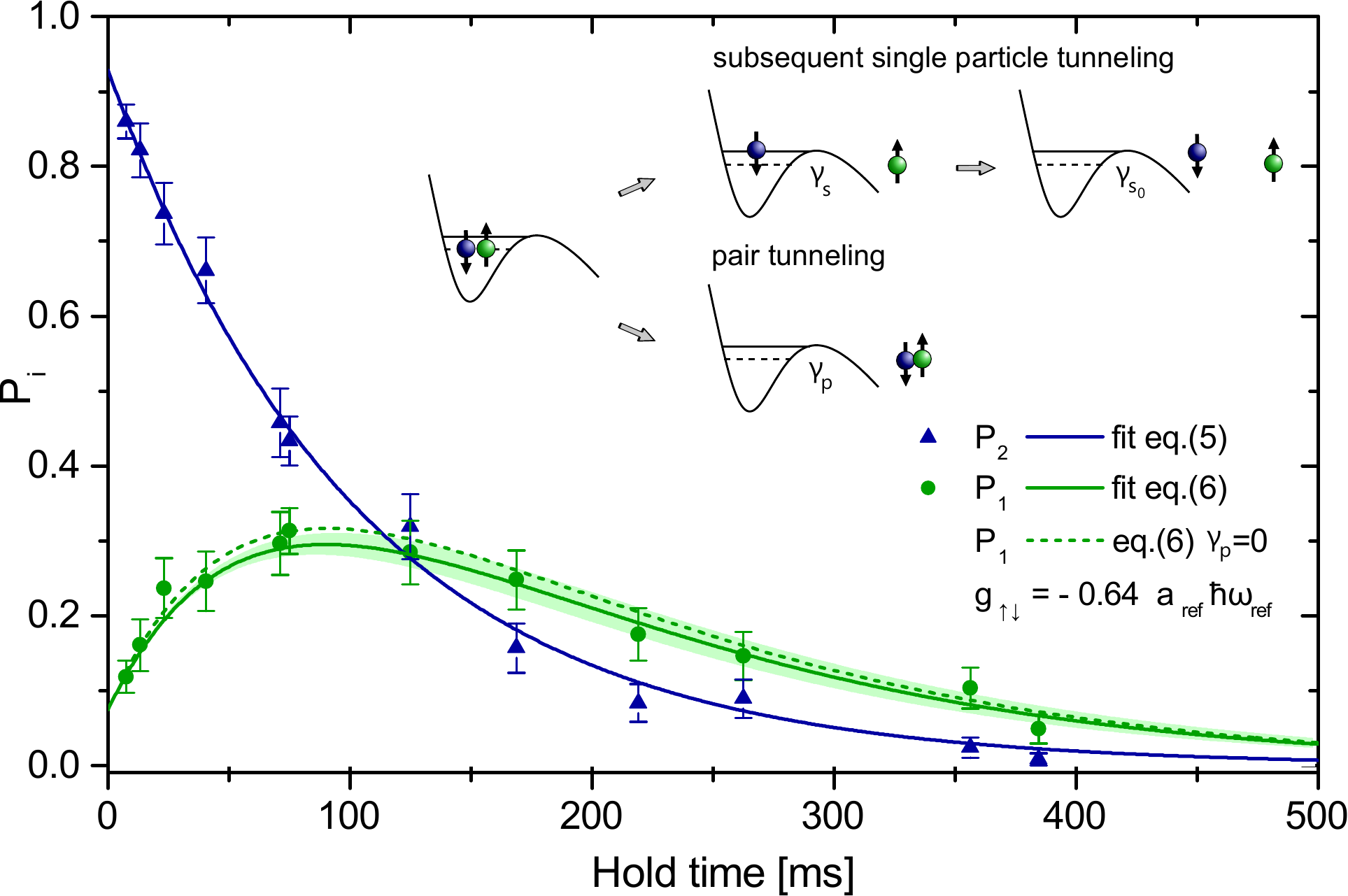}
	\caption{Measured time evolution of the probabilities $P_1(t)$ and $P_2(t)$ to find one (green dots) or two (blue triangles) particles in the tilted potential at an intermediate interaction strength of $g_{\uparrow \downarrow} = -0.64$. The blue solid line shows a fit of eq. \ref{eq:P2_rate_equation_mod} to $P_2(t)$ with free parameters $f$ and $\gamma_{2}$. The green line shows the fit to $P_1(t)$ with the single free parameter $\gamma_{p}$, where the shaded region indicates the uncertainty which results from our determination of the shape of the trapping potential. For comparison, the dashed line shows the result from eq. (6) with $\gamma_{p}$ set to zero, which is also consistent with our data. The errors are the $68\%$ confidence interval of about $100$ individual measurements. 
The inset shows a sketch of the loss processes included in our tunneling model: Subsequent single-particle tunneling with rates $\gamma_{s}$ and $\gamma_{s_0}$ and direct pair tunneling with a rate $\gamma_{p}$.
}
	\label{fig:figure2}
\end{figure}
For $g_{\uparrow \downarrow} > -0.59$ we observe no pair tunneling. For $g_{\uparrow \downarrow} = -0.64$ we find a pair tunneling rate of \nobreak{${\gamma_p/\gamma_2=7(4)(10)(\text{stat.})(\text{sys.})\%}$}. 
Therefore our data is consistent with a model which only considers subsequent single-particle tunneling. For stronger interaction pair tunneling is expected to play a stronger role. However, in our measurements for $g_{\uparrow \downarrow} < -0.64$  the probability of finding a single particle in the trap is only a few percent which is as small as the errors and consequently we cannot resolve to which extent the two particles tunnel as a bound object.

To compare the tunneling dynamics at different interaction strengths which occur on timescales differing by almost two orders of magnitude we rescale the data onto a common axis by plotting $P_1(t)$ as a function of the mean particle number (see Fig.\,\ref{fig:figure3} left panel). 

\begin{figure} [b!]
\centering
	\includegraphics [width=8.7cm] {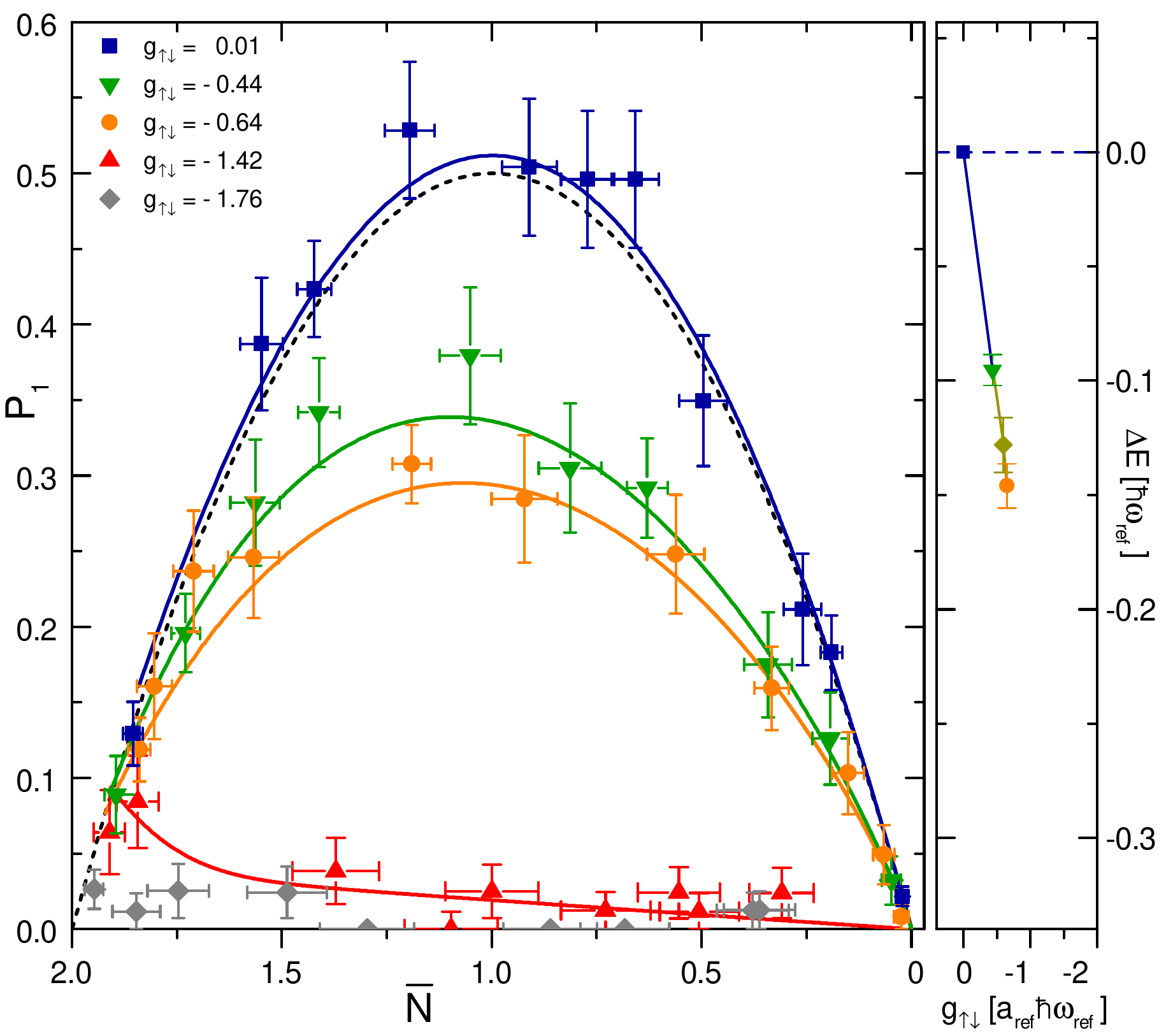}
	\caption{Probability of finding a single particle in the trap plotted as a function of the mean particle number $\overline{N}$. The solid lines show the results of fits to the data according to our tunneling model (see Fig.\,\ref{fig:figure2}). 
For a noninteracting system the probability follows the expectation value of completely uncorrelated tunneling given by the black dashed line. For increasing interaction strengths it becomes less likely to find a single particle in the trap, which we interpret as a result of increased pair-correlations. To quantify this effect we plot the interaction energy up to intermediate interaction strength as a function of $g_{\uparrow \downarrow}$ in the right panel. Here, pair tunneling is not considered in the analysis as it does not seem to play an important role at these interaction strengths.
}
	\label{fig:figure3}
\end{figure}
For a noninteracting system (blue) we find that the probability of finding one atom follows that of completely uncorrelated tunneling indicated by the black dashed parabola
$P_1  = \overline{N} - \overline{N}^2/2$.
For increasing attractive interaction the tunneling of the two particles is not uncorrelated anymore and we observe that the probability of finding a single particle in the trap decreases dramatically. This decrease in $P_1(t)$ can be explained by the fact that for larger interaction strengths the two-particle system experiences a larger effective barrier height and the tunneling rate $\gamma_{s}$ of the first particle decreases, while the tunneling rate ${\gamma_{s_0}}$ of the remaining particle is not affected. This leads to an decrease of the ratio ${\gamma_{s}}/{\gamma_{s_0}}$ and therefore a lower probability to observe a single particle in the trap, which is well described by our tunneling model. 

In the regime of weak interactions ($g > -0.64$) where pair tunneling plays only a negligible role we use our model to determine the amount of interaction energy that is released as the first particle tunnels from the trap,
which we call the separation energy.
This is done in an iterative process where we vary $ E_{\text{int}}$ and calculate $\gamma_{s\vert \uparrow \rangle}$ and $\gamma_{s\vert \downarrow \rangle}$ using a WKB calculation until the tunneling rates match the ones determined by a least squares fit to the $P_1$-data \cite{SOM}. 
One should note that our approach does not take into account the wavefunction overlap of the trapped state and the continuum state within the tunneling barrier. This leads to a systematic error in the interaction energy comparable to the one observed for repulsively interacting systems \cite{zuern2012, rontani2012a}
 which has recently been addressed in \cite{rontani2013}. 
However the qualitative behavior of the separation energy at fixed particle number or at fixed interaction strength is not expected to be altered by this systematic effect.

For a two-particle system the separation energy corresponds to the full interaction energy of the system. We clearly observe an increase of this interaction energy as a function of coupling strength, which is plotted in the right panel of Fig.\,\ref{fig:figure3}. Since in a 1D system a decrease of the energy of the two-particle ground state leads to a collapse of the two-particle wavefunction this corresponds to an increase of the local pair correlation $g^{(2)}(0)$ \cite{Busch1998, frankearnold2003}. 
This energy measurement therefore directly shows the appearance of pairing in our attractively interacting two-particle system.

After having used the two-particle system to understand the tunneling dynamics and establishing a method to determine the separation energies of our system we can now study how the observed pairing affects larger systems. In nuclear physics such pairing is one of the key ingredients required to obtain a quantitative understanding of the stability and binding energies of nuclei. One of its most pronounced consequences is the odd-even effect: Systems with an even number of neutrons or protons have larger binding energies and increased stability against decay. To study this phenomenon in our model system we measure
the separation energy for two- to six-particle systems with an interaction strength of \nobreak{${g_{\uparrow \downarrow} \approx -0.6}$} \cite{SOM} and observe a clearly enhanced stability of systems with even particle number (see Fig.\,\ref{fig:figure4}).
 
\begin{figure} [tb!]
\centering
	\includegraphics [width=6.5	cm] {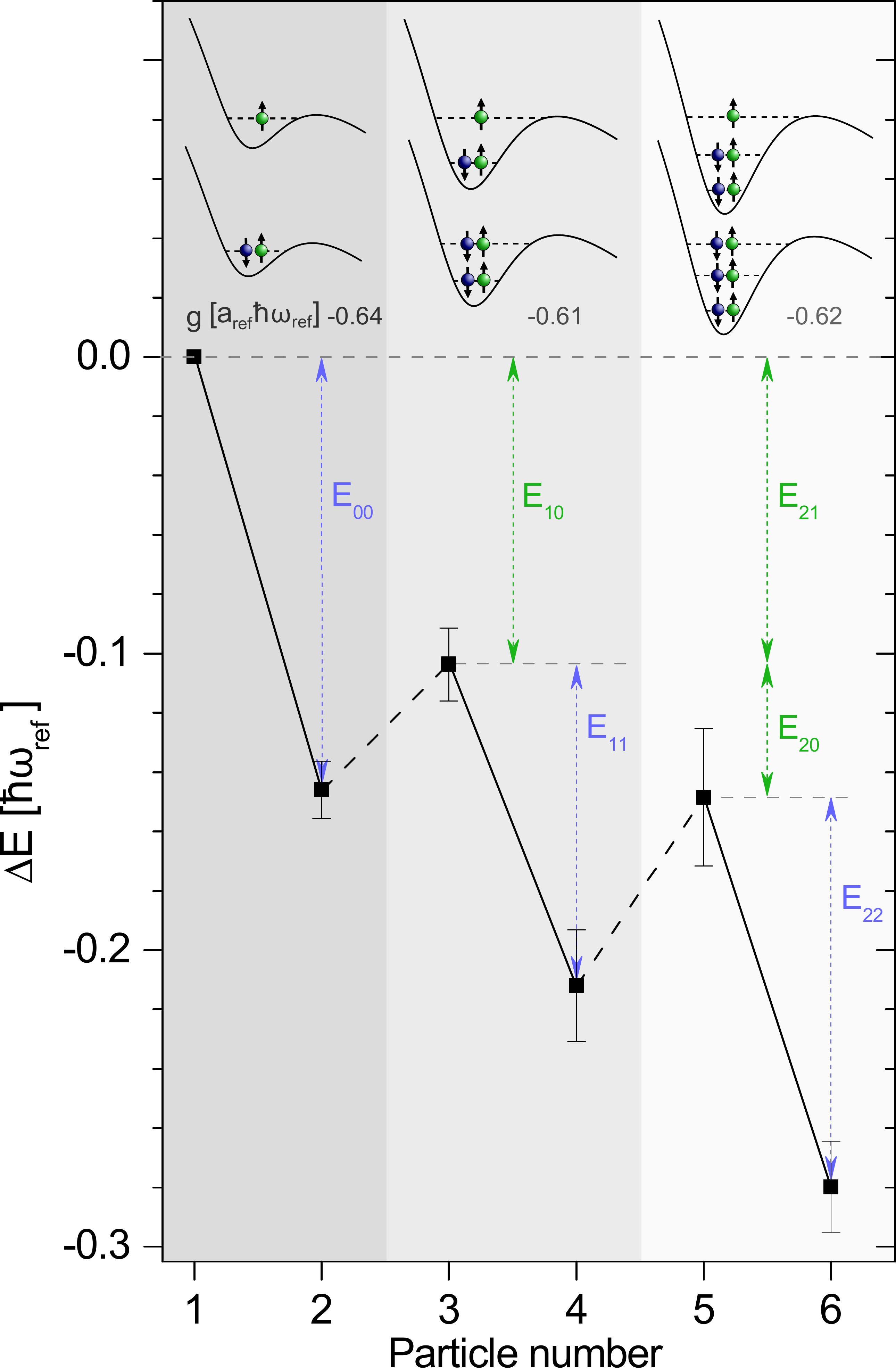}
	\caption{ 
	Separation energies for systems of $N=1\ldots6$ particles at an interaction strength of \nobreak{${g_{\uparrow \downarrow} \approx -0.6}$}. The small difference in $g_{\uparrow \downarrow}$ for the different particle numbers is due to the dependence of the coupling strength on the depth of the optical potential.
	The separation energies are normalized by the level spacing of the two uppermost trap states. The error bars show the relative uncertainty originating from the statistical errors of the fits of $\gamma_N$ and $\gamma_{s_0\vert \uparrow \rangle}$.
	The arrows indicate the contributions $E_{\text{ii}}$ and $E_{\text{ij}}$ of the intrashell (blue)  and intershell (green) interaction to the separation energy.
	}
	\label{fig:figure4}
\end{figure}

In addition to this odd-even effect we also observe a general decrease of the separation energy with growing particle number.
To quantify this effect we consider the single-particle levels in the trap as the shells of our 1D system and use the difference in the separation energy between open and closed-shell systems to estimate the contributions of intra- and intershell pairing.  For the two-particle system the separation energy $\Delta E$ directly corresponds to the pairing energy $E_{00}$ of the two particles on the lowest shell.
For $N=3$ the third particle has no interaction partner on its own shell. However, its separation energy is still strongly reduced compared to the noninteracting case. This suggests a strong intershell pairing ($E_{10}$) to the particles on the shell below. For $N=4$ the separation energy is then given by the sum of the intershell interaction energy $E_{10}$ and the intrashell interaction energy $E_{11}$ which are of similar size.

To understand the observed scaling of the separation energy with particle number we first consider a weakly interacting system. In this case two particles on the same shell obtain an interaction shift proportional to the coupling constant as they have the same spatial wavefunction.
To first order particles on closed shell have the same wavefunction and therefore Pauli blocking prevents all other particles on different shells to have any overlap with them.
Only in a second order process where the wavefunctions of the particles in the closed shell are modified due to the interactions with a  particle on a different shell the overlap becomes non-zero.
From the fact that the observed intershell pairing energy is comparable to the intrashell pairing energy we conclude that the different shells have a significant overlap and our system is no longer in the weakly interacting regime. 
Therefore our results can serve as a test for theories which predict the disappearance of shell structures in strongly interacting few-body systems \cite{Rontani2009, Zinner2009}.

In conclusion, we have used tunneling experiments to study quasi-1D few-fermion systems with attractive interactions and observed the emergence of correlations in the tunneling dynamics. We have developed 
a model which accurately describes the tunneling dynamics of the two-particle system and used it to infer the presence of pair correlations. 
We have then used this model to determine the separation energies for larger systems and identified the contributions of intra- and intershell pairing to the observed odd-even effect.
These measurements open the door to study how the shell structure of a finite system evolves in the BEC-BCS crossover \cite{Rontani2009, Zinner2009}. This would also be the first step towards studying the emergence of BCS-like superfluidity in a finite system of ultracold atoms through studies of rotational excitations \cite{Migdal1959, Grebenev1998}.\\

We thank Massimo Rontani for inspiring discussions and valuable theoretical input. 
We thank Shannon Withlock for helpful discussions.
The authors gratefully acknowledge support from ERC starting grant $279697$, the Helmholtz Alliance HA$216/$EMMI and the Heidelberg Center for Quantum Dynamics.  G.Z. and A.N.W. acknowledge support by the IMPRS-QD.

\color[rgb]{1,1,1}

\begin{equation}
\label{eq:dummy}
\end{equation}

\color[rgb]{0,0,0}

\cleardoublepage

\section*{SUPPLEMENTAL MATERIAL}

\subsection*{The confining potential}
The confining potential is created by a cigar-shaped optical dipole trap created by the focus of a single laser beam. In a harmonic approximation the trap frequencies of the optical potential are $\omega_{\parallel}=2 \pi \times 1.488(14)$\,kHz \cite{zuern2012} in longitudinal and $\omega_{\perp}=2 \pi \times 14.22(35)$\,kHz \cite{sala2013} in perpendicular direction. To realize a finite potential barrier through which the atoms can tunnel out of the trap we additionally apply a magnetic field gradient in longitudinal direction. This creates a magnetic potential which depends on the magnetic moment of the atoms and on the strength of the magnetic field gradient. 
For our calculations we parametrize the potential by 
\begin{equation}
V_{r=0}(z)= pV_0(1- \frac{1}{1+(z/z_R)^2})- c_{B\vert\text{state}\rangle} \mu_B \,B' z,
\label{eq:spindependentpotential}
\end{equation}
where $V_0= k_B \,3.326 \,\mu$K  is the initial depth of the optical potential, $p$ is the optical trap depth in units of $V_0$, $z_R= \frac{\pi \, w_0^2}{\lambda}$ is the Rayleigh range of the optical trapping beam with focal waist $w_0= 1.838 \, \mu$m and wavelength $\lambda=1064\,$nm, $\mu_B$ is the Bohr magneton, $c_{B\vert\text{state}\rangle}$ is a state and field dependent coefficient (see next section) and $B'= 18.92$ G/cm is the strength of the magnetic field gradient. The determination of the trap parameters is described in \cite{zuern2012}.
To modify the timescale on which the particles tunnel through the potential barrier we vary the trap depth of the optical potential which is quantified by the trap depth parameter $p$.

\subsection*{The state and magnetic field dependence of the potential}

For our experiment we use $^6\text{Li}$ atoms in the $F=1/2$, $m_F=1/2 $ and $F=3/2$, $m_F=-3/2$ hyperfine states which we label $\mid \uparrow \rangle$ and $\mid \downarrow \rangle$. The magnetic moment of these states depends on the magnetic offset field as shown in Fig.\,\ref{fig:SOM1}a) \cite{BreitRabi1931}, which results in a state and magnetic field dependent magnetic potential $V_{\text{mag}}=c_{B\vert\text{state}\rangle} \mu_B B' z$. 
\begin{figure} [th!]
\centering
	\includegraphics  [width=8.7cm] {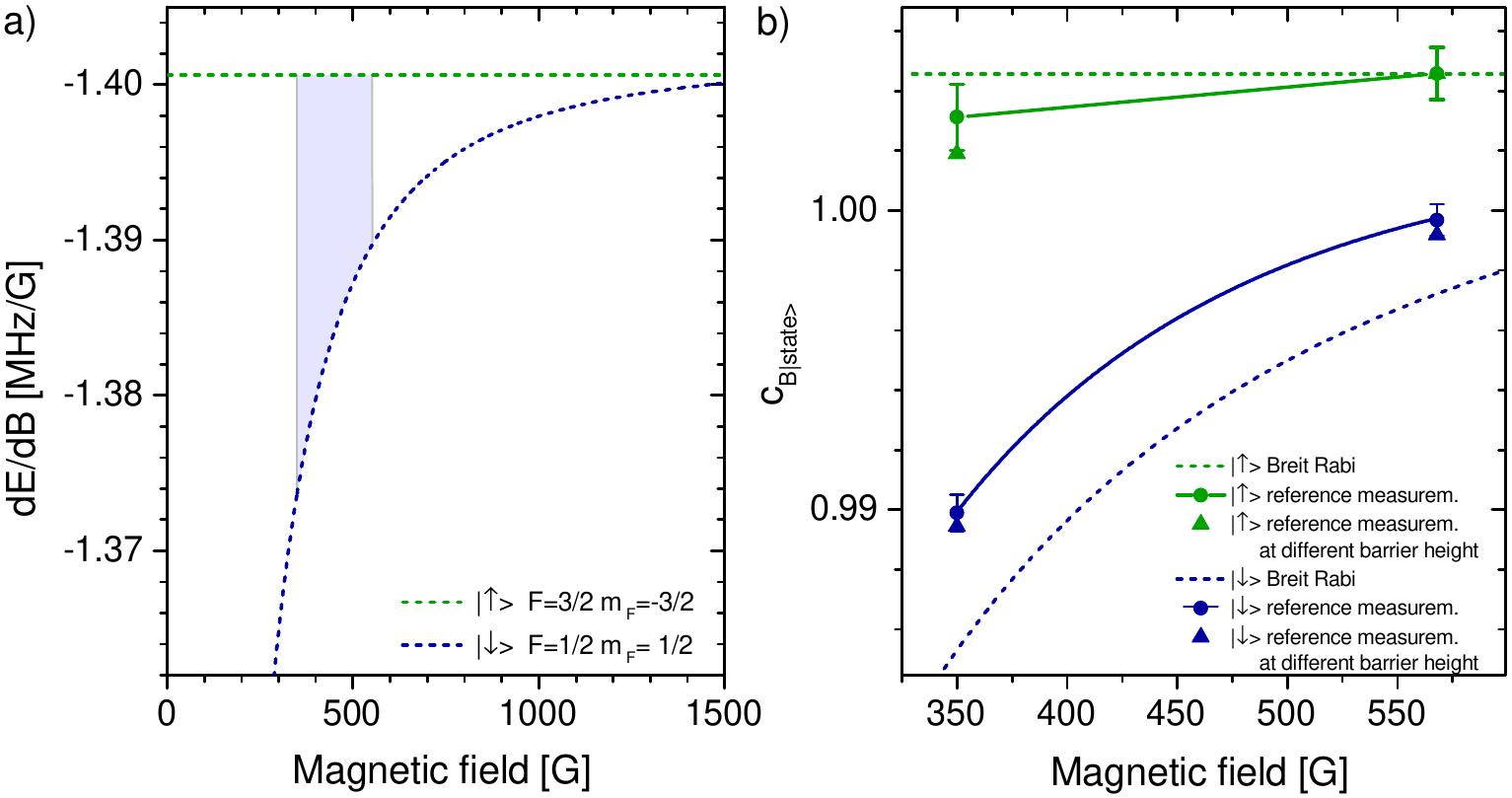}
	\caption{	a) Magnetic field dependence of the magnetic moments of $^6$Li atoms in the two hyperfine states used in our experiments. In the region of magnetic offset fields where we perform the tunneling measurement the magnetic moments differ by more than $1\%$ (blue shaded area). 
b) Coefficients $c_{B\vert\text{state}\rangle}$ according to eq.\,(\ref{eq:spindependentpotential}), which we determine for states $\mid \uparrow \rangle$ (green) and $\mid \downarrow \rangle$ (blue) at different magnetic fields and for two different barrier heights (dots and triangles). The dashed lines show the prediction of the Breit Rabi formula \cite{BreitRabi1931}; the solid lines show the interpolation used to determine $c_{B\vert\text{state}\rangle}$  for intermediate magnetic fields.
		}
	\label{fig:SOM1}
\end{figure}
To describe our tunneling experiment we therefore have to determine the shape of the potential for both states and all used magnetic offset fields. For this we determine the coefficient $c_{B\vert\text{state}\rangle}$ by performing reference measurements with single $\mid \uparrow \rangle$ or $\mid \downarrow \rangle$-particles for different offset fields at the same optical power and magnetic field gradient. The measured single particle tunneling rates are given in table 	\ref{tab:magnetic_moment_coefficient}.
\bgroup
\def\arraystretch{1.2}%
\begin{table}[t]
\begin{center}
  \begin{tabular}{|c | c |c|c| c| c| }
    \hline
magnetic field $B$ & $\vert \text{state}\rangle$ 	&   $\gamma_{B \vert \text{state}\rangle}$	& $\sigma_{\gamma_{B\vert \text{state}\rangle}}$ & $c_{B\vert\text{state}\rangle}$ & $\sigma_{c_{B\vert\text{state}\rangle}}$ \\
$[$G$]$								& 											&    [1/s] 								& [1/s] &    & [$10^{-4}$] \\ \hline
569	&$\mid \uparrow \rangle$	  &35.25	&3.57 & 1.00457 & 8.7\\
350	&$\mid \uparrow \rangle$	  &30.12	&2.81 & 1.00311 & 11.0\\ 
569	&$\mid \downarrow \rangle$	&21.76	&1.12 & 0.99968 & 5.3\\
350	&$\mid \downarrow \rangle$	&\,8.28	&0.49 & 0.98989 & 6.0\\ \hline
496	&$\mid \uparrow \rangle$	  & \multicolumn{2}{c|}{\footnotesize{interpol.}}  & 1.00407 & 11 \\ 
423	&$\mid \uparrow \rangle$    & \multicolumn{2}{c|}{\footnotesize{interpol.}}  & 1.00356 & 11 \\ 
496	&$\mid \downarrow \rangle$	& \multicolumn{2}{c|}{\footnotesize{interpol.}}  & 0.99806 & 6 \\ 
423	&$\mid \downarrow \rangle$	& \multicolumn{2}{c|}{\footnotesize{interpol.}}  & 0.99512 & 6 \\ \hline
$>$850 \; & $\mid \uparrow \rangle \mid \downarrow \rangle$ &\multicolumn{2}{c|}{$c_{\text{569G}\vert \uparrow \rangle}$}  &1.00457& 8.7\\ \hline
     \end{tabular}
\end{center}
\caption{Potential coefficients $c_{B\vert\text{state}\rangle}$ for the two states at different magnetic fields as extracted from the tunneling rates of $\gamma_{B \vert \text{state}\rangle}$ observed in the reference measurements. For magnetic field values between $350\,$G and $569\,$G we use an interpolation indicated by the solid lines in Fig.\,\ref{fig:SOM1}b). Above $850\,$G the relative difference in magnetic moment for the different states is smaller than $3 \times 10^{-3}$ and thus we set $c_{>\text{850G}\vert \text{state} \rangle}=c_{\text{569G}\vert \uparrow \rangle}$. \\
}
	\label{tab:magnetic_moment_coefficient}
\end{table}
\begin{figure} [th!]
\centering
	\includegraphics  [width=8.7cm] {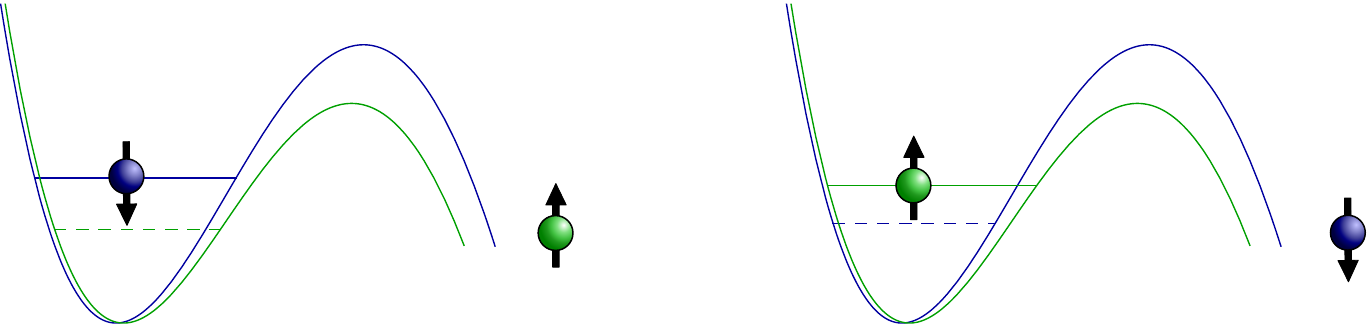}
	\caption{Sketch of the state dependent potential for the different hyperfine states at a magnetic offset field of $350\,$G. The green curve shows the shape of the potential for an $\mid \uparrow \rangle$-particle, the blue curve the one for a $\mid \downarrow \rangle$-particle. The two sketches illustrate the two different possibilities of subsequent single particle tunneling with either an $\mid \uparrow \rangle$-particle tunneling first (left) or a $\mid \downarrow \rangle$-particle tunneling first (right). The dashed lines indicate the energies of the particles which have tunneled first.}
	\label{fig:som2}
\end{figure}
From these measurements we then deduce the form of the state dependent potential by using the same approach as employed in ref. \cite{zuern2012}. We match the result of a WKB calculation for the trial potential given by eq.(\ref{eq:spindependentpotential}) to the experimentally determined field and state dependent tunneling rates where $p$ and $c_{B\vert\text{state}\rangle}$ are the only free parameters.
As we use the same value of $B'$ for our parametrization of the potential as in \cite{zuern2012} the coefficient $c_{\text{792G}\vert \downarrow \rangle}$ is $1$. From the functional form of the magnetic moment plotted in Fig.\,\ref{fig:SOM1}a) we then calculate $c_{\text{569G}\vert \uparrow \rangle}$ to be 1.00407. 
Using the WKB calculation we can now fix the optical potential parameter $p_{\text{ref}}$ of the reference measurement at $569\,$G by matching the calculated tunneling rate  of an $\mid \uparrow \rangle$-particle to the experimental one. 
Having fixed the optical potential at $p_{\text{ref}}$ we can finally determine $c_{\text{350G}\vert \uparrow \rangle}$ , $c_{\text{569G}\vert \downarrow \rangle}$  and $c_{\text{350G}\vert \downarrow \rangle}$ by matching the tunneling times from the WKB calculation with the ones  measured at the corresponding magnetic offset field.
The results are shown in table \ref{tab:magnetic_moment_coefficient} and Fig.\,\ref{fig:SOM1}b) and the shape of the determined potential is illustrated in Fig.\,\ref{fig:som2}.
To check if this approach is robust we have repeated the reference measurement for a larger optical trap depth. This increase in the barrier height led to a decrease of the tunneling rates by a factor of $5$ but did not lead to a significant change of the coefficients $c_{B\vert\text{state}\rangle}$.
However, if we compare our results to theory we observe a slight deviation from the prediction given by the Breit Rabi formula (see Fig.\,\ref{fig:SOM1}b)). This might be due to an additional magnetic field gradient created by the offset coils or due to a systematic error resulting from a difference between the shape of our model potential and the actual trapping potential.
However, as in our experiment the relevant information is obtained by comparing different tunneling rates where these systematic errors cancel to first order, we neglect this small systematic error in the analysis of our tunneling data.

\subsection*{Determination of the noninteracting single particle tunneling rates }

For the fit of our tunneling model we have to determine the tunneling rates $\gamma_{s_0\vert\uparrow\rangle}$ and $\gamma_{s_0\vert\downarrow\rangle}$ of noninteracting $\mid \uparrow \rangle$ and $\mid \downarrow \rangle$-particles. These rates depend only on the previously determined coefficients $c_{B\vert\text{state}\rangle}$ and on the trap depth parameter $p$ which was the same for all measurements with same particle number.
To determine the value of $p$ we use our tunneling measurement at $569\,$G where $g_{\uparrow \downarrow}=0$ and the two noninteracting particles follow a double-exponential decay of the form
\begin{equation}
	\overline{N}=N_0\, \left(e^{-\gamma_{s_0,569\text{G}\vert \uparrow \rangle}t}+ e^{-\gamma_{s_0,569\text{G}\vert \downarrow \rangle}t} \right).
\label{eq:douple_exp_SOM1}
\end{equation}
For our fit we rewrite this equation as 
\begin{equation}
	\overline{N}=N_0\, \left(e^{-\gamma_{s_0,569\text{G}\vert \uparrow \rangle}t}+ e^{-r\,\gamma_{s_0,569\text{G}\vert \uparrow \rangle}t} \right)
\label{eq:douple_exp_SOM}
\end{equation}
with the parameter $r=\gamma_{s_0,569\text{G}\vert \downarrow \rangle}/\gamma_{s_0,569\text{G}\vert \uparrow \rangle}<1$ which we iteratively fit to the measured data.
For each iteration we start by fitting eq.(\ref{eq:douple_exp_SOM}) to the data with a fixed value of $r$ and $\gamma_{s_0,569\text{G}\vert \uparrow \rangle}$ as a free parameter. Then we perform a WKB calculation where we tune $p$ such that $\gamma_{s_0,569\text{G}\vert \uparrow \rangle,\text{WKB}}$ is identical to $\gamma_{s_0,569\text{G}\vert \uparrow \rangle,\text{fit}}$.
The parameter $r$ for the next step of the iteration is determined by calculating $\gamma_{s_0,569\text{G}\vert \downarrow \rangle,\text{WKB}}\, /\gamma_{s_0,569G\vert \uparrow \rangle,\text{WKB}}$.
We stop the iteration when $\gamma_{s_0,569\text{G}\vert \text{state} \rangle, \text{fit}}$ and $\gamma_{s_0,569\text{G}\vert \text{state} \rangle,\text{WKB}}$
coincide with a precision of $10^{-3}$. The derived optical trap depth parameters $p$ which fix the final free parameter of the potential are given in table \ref{tab:optical_power_attractive}.
From this we can calculate the single particle tunneling rates for different magnetic offset fields which are given in table III.
\begin{table}[h!]
\begin{center}
  \begin{tabular}{|c | c | c |   }
    \hline
    prepared particle  & potential parameter $p$ & $\omega_{\text{ref}}$ \\ 
    number $N$ & [fraction of initial depth] & [Hz] \\ \hline
		 2 & 0.63496 & 640$\,\pm\,$19 \\ \hline
     3 & 0.69232 & 762$\,\pm\,$13 \\ 
		 4 & 0.69232 & 762$\,\pm\,$13 \\ \hline
		 5 & 0.73227 & 778$\,\pm\,$12 \\ 
		 6 & 0.73136 & 778$\,\pm\,$12 \\ \hline
     \end{tabular}
\end{center}
\caption{Optical trap depth parameter $p$ for the different N-particle systems and corresponding reference frequency $\omega_{\text{ref}}$.}
	\label{tab:optical_power_attractive}
\end{table}

\begin{table*}[t!]
\begin{minipage}[b]{17.7cm}
\begin{centering}
  \begin{tabular}{|c|c |c  |c | c |c|c|c|  }
    \hline
		$B$&$g_{\uparrow \downarrow}$	& $\gamma_{s_0\vert \downarrow \rangle,\text{WKB}}$  & $\gamma_{s_0\vert \uparrow \rangle,\text{WKB}}$   & $\gamma_{2,\text{fit}}$ 	&  $\gamma_{s\vert \downarrow \rangle,\text{WKB}}$ 		& $\gamma_{s\vert \uparrow \rangle,\text{WKB}}$ & $E_{\text{int,WKB}}$  \\
$[$G$]$&$[a_{\text{ref}}\hbar\omega_{\text{ref}}]$ & [1/s] &   [1/s]  &  [1/s]   &  [1/s]   & [1/s]    &   $[\hbar\omega_{\text{ref}}]$ \\ \hline
496&-0.44 &  21.1 & 36.3  & 22.2$\,\pm\,$1.0			&	8.05		&		14.12		&	-0.095$\,\pm\,$0.007 	\\
423&-0.59 &  15.85& 34.9  & 13.84$\,\pm\,$1.04		  &	4.23		&		9.62	& -0.128$\,\pm\,$0.012	\\
350&-0.64 &  9.44 & 33.7  & 9.70$\,\pm\,$0.33		  &	2.03		&		7.67		&	-0.146$\,\pm\,$0.010 	\\ \cline{3-4} \cline{6-7}
1202&-1.42&  	\multicolumn{2}{c|}{\footnotesize{$\gamma_{s_0 \vert \uparrow \rangle,\text{fit}}$}}    & 2.14$\, \pm\,$0.19		  &	\multicolumn{2}{c|}{\footnotesize{$\gamma_2 /2$}}  	&	(-0.320$\,\pm\,$0.034)			\\ 
1074&-1.48&  	\multicolumn{2}{c|}{\footnotesize{$\gamma_{s_0 \vert \uparrow \rangle,\text{fit}}$}}    & 1.931$\,\pm\,$0.123		&	\multicolumn{2}{c|}{\footnotesize{$\gamma_2 /2$}} 		&	(-0.327$\,\pm\,$0.028)\\
958&-1.57 &  	\multicolumn{2}{c|}{\footnotesize{$\gamma_{s_0 \vert \uparrow \rangle,\text{fit}}$}}    & 1.227$\,\pm\,$0.053		&	\multicolumn{2}{c|}{\footnotesize{$\gamma_2 /2$}} 		&	(-0.357$\,\pm\,$0.025)			\\
851&-1.76 &  	\multicolumn{2}{c|}{\footnotesize{$\gamma_{s_0 \vert \uparrow \rangle,\text{fit}}$}}    & 0.505$\,\pm\,$0.023		&	\multicolumn{2}{c|}{\footnotesize{$\gamma_2 /2$}} 		&	(-0.408$\,\pm\,$0.030)			\\ \hline
     \end{tabular}
\end{centering}
	\end{minipage}
	\begin{minipage}[b]{17.7cm} 
	\begin{itemize}
		\item[]
		\item[]{TABLE III: Tunneling rates of a two-particle system for different coupling strengths. For $g_{\uparrow \downarrow}<-0.64$ we cannot determine to which extend the particles tunnel subsequently as the probability of finding one particle in the trap tends to zero. Hence the interaction energies determined under these conditions have large systematic errors and are therefore given in parentheses.}
	\label{tab:tunneling rates 2-particle system}
  \end{itemize}
	\hfill
	\end{minipage}
\begin{minipage}[b]{17.7cm}
\begin{centering}
\footnotesize
  \begin{tabular}{|l|r | c |c|c|c|c|c| c|  }
    \hline
\multicolumn{2}{|l|}{N} 	 & $\gamma_{s_0 \vert \downarrow \rangle,\text{fit}}$ 	&  $\gamma_{s_0 \vert \uparrow \rangle,\text{fit}}$ & $g_{\uparrow \downarrow}$	  & 		 $\gamma_{N,\text{fit}}$ & $\gamma_{s \vert \downarrow \rangle,\text{WKB}}$ & $\gamma_{s \vert \uparrow \rangle,\text{WKB}}$ & $E_{\text{diss,WKB}}$  \\
\multicolumn{2}{|l|}{} &  [1/s]   &  [1/s] & $[a_{\text{ref}}\hbar\omega_{\text{ref}}]$ & [1/s]  & [1/s] & [1/s]  &   $[\hbar\omega_{\text{ref}}]$ \\ \hline
\multicolumn{2}{|l|}{2 g=0}  & 24.8$\,\pm\,$1.4	&	  37.7$\,\pm\,$2.1		&      	&                 &     	&	        	&              	     \\         
\multicolumn{2}{|l|}{\,\tiny{g=-0.64}}  &  \; 9.44\,\scriptsize(WKB)	 \;	&	 \;  33.7\,\scriptsize(WKB)   \;     		&-0.64	&9.70$\,\pm\,$0.33& 2.03	&	7.67			&	-0.146$\,\pm\,$0.010\\ \hline
\multicolumn{2}{|l|}{3}  &			-						&		9.83$\,\pm\,$1.03		&				&3.43$\,\pm\,$0.19&	-			&	$\gamma_{3,\text{fit}}$&	-0.104$\,\pm\,$0.012\\ \cline{1-4} \cline{6-9}
\multicolumn{2}{|l|}{4 g=0}& 5.51$\,\pm\,$0.45&		9.78$\,\pm\,$0.79		&-0.61 & &				&						&				\\
\multicolumn{2}{|r|}{\,\tiny{g=-0.61}}  &	0.67\,\scriptsize(WKB)						&		$\gamma_{3,\text{fit}}$	 				& &1.59$\,\pm\,$0.06													&	0.25	&	1.34			&	-0.212$\,\pm\,$0.019\\ \hline
\multicolumn{2}{|l|}{5}  &			-						&		12.15$\,\pm\,$1.85	&				&3.09$\,\pm\,$0.11&	-			&	$\gamma_{5,\text{fit}}$&	-0.148$\,\pm\,$0.023\\ \cline{1-4} \cline{6-9}
\multicolumn{2}{|l|}{6 g=0} &7.79$\,\pm\,$0.29	&		14.32$\,\pm\,$0.53	&-0.62	&									&				&						&	     	\\
\multicolumn{2}{|r|}{\,\tiny{g=-0.62}}   &	0.54\,\scriptsize(WKB)					&		$\gamma_{5,\text{fit}}$	   			&				&1.49$\,\pm\,$0.06&	0.22	&	1.28			&	-0.280$\,\pm\,$0.015\\ \hline
     \end{tabular}
\end{centering}
	\end{minipage}
	\begin{minipage}[b]{17.7cm} 
	\begin{itemize}
		\item[]
		\item[]{TABLE IV: Tunneling rates and separation energies determined for systems with two to six particles. 
		Here  $\gamma_{s_0 \vert \uparrow \rangle}$ ($\gamma_{s_0 \vert \downarrow \rangle}$) is the tunneling rate of a single particle tunneling from the uppermost shell which is not interacting with another particle on the same shell. $\gamma_{N}$ denotes the rate with which the probability of finding $N$ particles in the trap decreases. $\gamma_{s \vert \uparrow \rangle}$ ($\gamma_{s \vert \downarrow \rangle}$) is the tunneling rate of the particle that tunnels first from an $N$-particle system.
		}
  \end{itemize}
	\hfill
	\end{minipage}
\end{table*}

\subsection*{Energy scale of the system and coupling constant	}
To compare our measurements performed with different particle numbers and therefore different trap depths we need to rescale the coupling constant and the measured interaction energies with the natural length and energy scales of the system. For a harmonic trapping potential these are the level spacing $\hbar \omega_\parallel$ and the harmonic oscillator length $a_\parallel= \sqrt{\hbar/ \mu \omega_\parallel}$. 
For our tilted potential we define the characteristic energy scale $\hbar \omega_{\text{ref}}$  as the level spacing between the two least bound states in the potential. If there is only one bound state left in the potential we define  $\hbar \omega_{\text{ref}}$ as twice the zero point energy $E_0$ of this ground state. We calculate the energies $E_i$ of these bound states with a WKB calculation using the potential parameters given in table 	\ref{tab:magnetic_moment_coefficient} and 	\ref{tab:optical_power_attractive}.
To determine a common reference at a fixed trap depth but for different states and offset fields, we average the frequencies $\omega_{\text{ref}\vert\text{state}\rangle \text{field}}$ of state $\mid\uparrow\rangle$ and $\mid\downarrow\rangle$ at $350\,$G and $569\,$G. The resulting reference frequencies are listed in table \ref{tab:optical_power_attractive}.\\
The strength of the interparticle interaction is given by the 1D coupling constant 
\begin{equation}
g_{\uparrow \downarrow}=\frac{2\hbar^2a_{\text{3D}}}{\mu a_{\perp}^2}\frac{1}{1-Ca_{\text{3D}}/a_{\perp}},
\label{eq:olshanii}
\end{equation}
which can be calculated from the 3D scattering length $a_{\text{3D}}$  and the harmonic oscillator length $a_{\perp}=\sqrt{\hbar/ \mu \omega_{\perp}}$ in the perpendicular direction \cite{Olshanii1998}, where $\hbar$ is the reduced Planck constant, $\mu=\frac{m}{2}$ the reduced mass of two $^6$Li atoms with mass $m$ and $C=-\zeta(\frac{1}{2})\approx 1.46$ where $\zeta(x)$ is the Riemann zeta function. 
This coupling constant can be tuned over a wide range by changing the value of the 3D scattering length $a_{\text{3D}}$ using a magnetic Feshbach resonance. 
To determine the coupling constant  of our system we evaluate eq.(\ref{eq:olshanii}) by using the scattering length of $^6$Li \cite{zuern2013} and the harmonic confinement length $a_\perp(p)=\sqrt[4]{p}a_{\perp}$. We give $g_{\uparrow \downarrow}$ in units of our reference frequency $\omega_{\text{ref}}$ and of $a_\text{ref}= \sqrt{\hbar/ 
\mu \omega_\text{ref}}$. 
Note that for two identical fermions s-wave scattering is forbidden and thus $g_{\uparrow \uparrow}$ = 0 and $g_{\downarrow \downarrow}$ = 0.

\subsection*{Fitting procedure for the tunneling model with attractive interaction}

To fit our tunneling model to the measured data we use a combination of least squares fits and WKB calculations.
As the noninteracting single particle decay rates $\gamma_{s_0 \vert \uparrow \rangle}$ and $\gamma_{s_0 \vert \downarrow \rangle}$ are fixed by the shape of the potential there are three free parameters ($\gamma_{s\vert \uparrow \rangle}$, $\gamma_{s \vert \downarrow \rangle}$ and $\gamma_{p}$) left in the model which we use to describe the tunneling process (eq.\,(6)). 
However, there is one additional condition which links $\gamma_{s\vert \uparrow \rangle}$ and  $\gamma_{s\vert \downarrow \rangle}$:
In the case that a single particle tunnels out of the trap the remaining particle is left in the unperturbed ground state of the potential. Hence the full interaction energy is given to the tunneled particle.
As we know the shape of the potential we can calculate its tunneling rate $\gamma_{s\vert \text{state}\rangle}$ from its energy $E=E_0-E_{\text{int}}$ using a WKB calculation.
With this we obtain a relation for $E_{\text{int}}$  which monotonically increases with the inverse tunneling rate $\gamma_{s\vert \text{state}\rangle}^{-1}$. 
Hence in an iterative process we can vary $ E_{\text{int}}$ and calculate $\gamma_{s\vert \uparrow \rangle}$ and $\gamma_{s\vert \downarrow \rangle}$ with the WKB calculation until  $\gamma_{s\vert \uparrow \rangle}+\gamma_{s\vert \downarrow \rangle}+\gamma_p=\gamma_2$, where $\gamma_p$ is determined by a least squares fit to the $P_1$-data in each step of the variation.\\
The resulting energies are given in table III where we have set $\gamma_p=0$, as we have observed that pair tunneling plays only a negligible role for $g_{\uparrow \downarrow}\geq-0.64$. \\
To determine the separation energy of systems with $N=2\ldots6$ particles \cite{GerhardPhD} we use a modified version of this procedure:
Due to the large level spacing of our microtrap we can set the barrier height such that only the particles in the uppermost level can tunnel out of the trap on the timescale of the experiment. This allows us to restrict our tunneling model to these uppermost particles and, hence, we can apply the model established for two particles also to larger systems.
For odd particle numbers we fit a single-particle decay of the particle highest up in the potential and extract the separation energy from the decay rate. 
In the case of a system with even particle number the single particle which remains on the uppermost shell still interacts with the particles occupying lower shells. 
Hence when applying the model of subsequent single-particle tunneling, $\gamma_{s_0,\vert\text{state}\rangle}$ is not the rate of a noninteracting particle, but of a single particle interacting with two or four other particles in lower shells. 
For $\gamma_{s_0,\vert \uparrow \rangle}$ we determine this rate directly by measuring the single-particle decay rate of a three and a five-particle system (see table IV).
From this measured rate we can also determine $\gamma_{s_0,\vert \downarrow \rangle}$ of a four and a six-particle system by using a WKB calculation.\\
To account for the finite lifetime of the particles on the lower shells  we add an additional decay term $e^{N\gamma_{l2(4)}}$to the probability to find $N$ particles in the trap, where we measured the $1/e$-decay rate of two (four) particles to be $\gamma_{l2}=0.0250 \pm 0.0004\;\;1/s$ $(\gamma_{l4}=0.035 \pm 0.002 \;\; 1/s)$ at a trap depth of $p=0.692$ $(p=0.731)$.

\footnotesize
	\begin{itemize}
		\item[]{
		\begin{center}
		\textbf{\_\_\_\_\_\_\_\_\_\_\_\_\_\_\_\_\_\_\_}
		\end{center}
		}
		\item[{[12]}]{\footnotesize G. Z\"urn et al.,   Phys. Rev. Lett. \textbf{110}, 135301, (2013).}
		\item[{[14]}]{\footnotesize G. Z\"urn et al.,   Phys. Rev. Lett. \textbf{108}, 075303, (2012).}
		\item[{[15]}]{\footnotesize S. Sala et al.,     Phys. Rev. Lett. \textbf{110}, 203202, (2013).}
		\item[{[18]}]{\footnotesize M. Olshanii,        Phys. Rev. Lett. \textbf{81}, 938, (1998).}
		\item[{[27]}]{\footnotesize G. Breit and I. I. Rabi, Phys. Rev. \textbf{38}, 2082 (1931).}
		\item[{[28]}]{\footnotesize G. Z\"urn, PhD thesis, Universit\"at Heidelberg (2012).}

  \end{itemize}
	\hfill


\begin{thebibliography}{28}
\bibitem[{\citenamefont{{In this work the two spin states $\mid \downarrow
  \rangle$ and $\mid \uparrow \rangle$ are realized by the $F=1/$2, $m_F=1/2 $
  and $F=3/2$, $m_F=-3/2$ hyperfine states of $^6$Li}}()}]{hyperfinestate}
\bibinfo{author}{\bibnamefont{{In this work the two spin states $\mid
  \downarrow \rangle$ and $\mid \uparrow \rangle$ are realized by the $F=1/$2,
  $m_F=1/2 $ and $F=3/2$, $m_F=-3/2$ hyperfine states of $^6$Li}}}.
\bibitem[{\citenamefont{{In the deformed potential $g_{\uparrow \downarrow}$ is
  given in units of $a_{\text{ref}} \hbar \omega_{\text{ref}}$
  \cite{SOM}}}()}]{unitg1d}
\bibinfo{author}{\bibnamefont{{In the deformed potential $g_{\uparrow
  \downarrow}$ is given in units of $a_{\text{ref}} \hbar \omega_{\text{ref}}$
  \cite{SOM}}}}.
\end{thebibliography}
\end{document}